Title: Colossal Van der Waals small-polaron superlattice: a hint to understanding the
   colossal magnetoresistance and electroresistance in manganites
Author: Mladen Georgiev (Institute of Solid State Physics, Bulgarian Academy of Sciences
   1784 Sofia, Bulgaria)
Comments: 9 wording pages with 3 (5) figures, all pdf format
Subject-class: physics

A mechanism is proposed to explain the formation of a small-polaron superlattice above the Curie temperature in manganites. The order-disorder transition initiated by the external field at a lattice is known responsible for the colossal resistance effect. We attribute the lattice formation to the occurrence of strong Van der Waals pairing of vibronic small polarons arising from the phonon coupling of highly polarizable two-level orbital systems. The latter having inherent electric and magnetic dipoles associated with them, they couple to the external field leading to the observed colossal effects. We find numerical estimates of the critical field strength compatible.

1. Rationale

The Colossal Magnetoresistance (CMR) and Colossal Electroresistance (CER) are amongst the thrilling discoveries of modern solid state physics made some 15 years ago [1]. Most of all they are observed as a multi-order-of-magnitude drop in electric resistance as a CMR- (CER-) active material is placed in an external magnetic (electric) field. The classical CMR (CER) material is manganese oxide (manganite) of the $La_{1-x}Sr_xMnO_3$ family, all of the perovskite $LaMnO_3$ structure in which a $La^{3+}$ rare earth ion is at the body center of a cube surrounded by 8 vertex $Mn^{3+}$ ions and 12 edge centered $O^{2-}$ ions, as shown in Figure 1(a) [2]. Figure 1(a) also depicts even parity vibrations of the oxygen octahedron around a $Mn^{3+}$ ion which can give rise to Jahn-Teller (JT) distortions ($E_g$ mode), possibly to JT-polarons. In the absence of an external field, the conductivity $\sigma$ of a CMR material is metallic below a conversion temperature $T_C$ (Curie) at which $\sigma$ is vanishing, the material turning insulating at $T \geq T_C$. It is believed that the conduction currents are carried by large polarons at $T \leq T_C$ which freeze into small polarons at $T \geq T_C$ that are far less mobile [3]. The small polarons form kind of a superlattice which is melted away by the external field bringing back the material to the original conductive (e.g. large-polaron) state. In effect, the applied external field displaces the conversion temperature $T_C$ towards the higher T along the temperatures axis. The superlattice melting has recently been proven by experiment, at least in its CER form [1].

The small polarons must each carry both individual electric and magnetic dipoles to couple to the external fields. In looking for a species, we remind of the off-center small polarons which are both electrostatically polarizable and carry orbital magnetic dipoles [4]. We had earlier proposed that dispersive pairing forces could arise within a field of vibronic small polarons to produce a binding mechanism. Now we shall pursue this matter in some more detail with regard to the colossal resistance effects in manganites.

The conversion mechanism is not yet clear though it suggests a field-dependent electron-phonon coupling constant G. In addition, G is strong though unconventional since one would normally expect small bound polarons to build up preceding the formation along the temperature axis of large polarons (else free carriers). One possible reason has been given in that the ferromagnetic (conductive) alignment of electron spins comes from an increased bandwidth which suppresses the small-polaron formation below the Curie temperature [3]. Nevertheless, the above mechanism may be expected to work better for Holstein's small polarons and less so for Jahn-Teller (JT)- or Pseudo-Jahn-Teller (PJT)- small polarons [4].

The free $Mn^{3+}$ ion 3d-electron quintet is split by the crystalline field into a higher-lying $e_g$ doublet and a lower-lying $t_{2g}$ triplet. If the $e_g$ doublet is singly occupied, it may couple to an even parity $E_g$ vibrational mode to produce a co-symmetric JT distortion of the surrounding $6O^{2-}$ oxygen octahedron, as in Figure 1(b). If not occupied, there will be no JT ($E_g$) splitting. Now a doublet level may mix vibronically through, say, an $T_{1u}$ odd-mode coupling with some odd-parity lower- lying level to produce a PJT off-center distortion of the $6O^{2-}$ octahedron at the Mn site. What is further needed is identifying the appropriate mixing partner.

On the other hand, one could hardly devise a simple magnetic field dependent electron-phonon coupling, which is the realm of electrostatics. $Mn^{3+}$ is a magnetic ion but, in addition, the Mn site may carry an orbital magnetic dipole if its $6O^{2-}$ oxygen frame is slightly displaced off center in a way shown in Figure 1(c) ($T_{1u}$ mode). As the off-center Mn site rotates jumping around the normal lattice site, the off-site dipole varies in both magnitude and direction and its average vanishes. Still, the off-site Mn is polarizable electrostatically and may give rise to a Van der Waals (VdW) coupling between neighboring Mn off-sites. The latter dispersive-type coupling may result in the formation of a periodic ferromagnetic or antiferromagnetic off-site orbital structure to host the small polarons. Another important conjecture is that $Mn^{3+}$ is a Jahn-Teller ion (cf. Figure 1(b)) leading to a vertical elongation of the oxygen octahedron (cf. Figure 1(a)). However, this distortion does not contribute to the Mn site electric dipole and thereby to the small-polaron superlattice or to the field coupling strength.

The small-polaron lattice, whether off-site or otherwise, seems to play an essential role in the field-dependent drop of electric resistance. Namely, the rise in polaron conductivity being heralded by the melting of the small-polaron lattice, the conductivity conversion at $T_C$ can also be interpreted as one from a solid polaron lattice to its liquid melt form (see Figure 2 for the potentials distinguishing small from large polarons). It is implied that the conductivity is hindered by the periodic small-polaron structure, which bounds the carriers, and is stimulated by this structure getting soft. The solid (small polarons) to liquid (large polarons) conversion is the heart of the CMR (CER) mechanism. All in all, lattice coupling effects have long been considered essential for the field-enhancement of electric conductivity of manganites [2].

Turning back to the small-polaron periodic structure, we stress that in our model it can be visualized as an array of consecutive left-hand and right-hand rotors kept intact by the dispersive VdW chemical bond. The dispersive force is enhanced by the lattice coupling reducing largely the electron gap energy, originally the order of an eV, to negligible values. This reduction called Holstein's band narrowing enhances the linear off-site polarizability and is the chief factor for the occurrence of a colossal VdW binding [4]. Now, if the sample with a small polaron superlattice, sustained by the colossal binding obliged to the enhanced linear polarizability, is placed in an electric field of sufficient strength, as it seems to be the case, the linear polarizability will be substituted for by the nonlinear or field-dependent polarizability which is considerably lower. In effect the polarizability lowering at the small-polaron sites would undoubtedly reduce the chemical binding therein causing the periodic structure to disintegrate.

The above being a plausible scenario for CER, the CMR mechanism requires additional consideration. At this point, we only suggest that the orbital magnetic dipoles of superlattice polarons should likely be aligned in the antiferromagnetic manner, by alternating left-hand and right-hand rotations. As an external magnetic field would tend to turn the dipole arrangements ferromagnetic, this would in effect increase the energy of the structure turning it less stable. In this way an external magnetic field of sufficient strength may again bring about the disintegration of the superlattice.

In what follows we shall pursue the colossal resistance problem along the above lines. We show that the picture outlined on this basis may be found compatible with the observations.

Before starting, it may be worth noting that CMR (CER) should not be confused with the "Giant Magnetoresistance" (GMR) observed in binary ferromagnetic layered structures which has a well documented spin-dependent mechanism [5]. Giant magnetoresistance (GMR) has first been observed in layered Fe/Cr magnetic structures. These structures have magnetized layers separated by a nonmagnetic spacer metal.

## 2. Field coupling to a small-polaron superlattice

### 2.1. Field-off analysis

We consider a molecular cluster with two nearly-degenerate opposite-parity electronic eigen-states $\psi_1$ and $\psi_2$ of eigen-energies $\varepsilon_1$ and $\varepsilon_2$, respectively. If now $\mathbf{p}_{12} = <\psi_2|\,e\mathbf{r}\,|\psi_1>$ is the electric dipole mixing these states, the electrostatic polarizability of the two-level system is

$$\alpha_{el} = (1/3)\, p_{12}^2 / |\varepsilon_2 - \varepsilon_1| \qquad (1)$$

The two-level electrostatic polarizability (1) brings about a VdW electronic pairing energy

$$U_{VdWel} = \tfrac{1}{2}\, \Delta E_{gap}\, (\alpha_{el} / \kappa\, R_{ij}^3)^2 = (p_{12}^4/18\kappa^2 E_{gap})(1/R_{ij}^6) \qquad (2)$$

$\kappa$ is the dielectric constant of the medium, $R_{ij}$ is the pair separation, $E_{gap} = |\varepsilon_2 - \varepsilon_1|$ is the interlevel energy gap. The pairing energy (2) arises from electrostatic interactions within the *bare two-level system*.

We subsequently assume that the two electronic states are coupled to mix *vibronically* by an odd-parity intermolecular vibration Q through the *pseudo-Jahn-Teller effect*. As a result of the vibronic mixing the original interlevel spacing $E_{gap}$ reduces to

$$\Delta E = E_{gap} \exp(-2E_{JT}/\eta\omega) \qquad (3)$$

(Holstein effect) where $E_{JT}=G^2/2K$ is the Jahn-Teller energy, $\omega \equiv \omega_{ren} = \omega_{bare}\sqrt{1 - (E_{gap}/4E_{JT})^2}$ is the renormalized frequency of the coupled vibration, $\omega_{bare}$ being the bare vibrational frequency, $\eta = h/2\pi$. The electrostatic polarizability of the *squeezed two-level system* turns into

$$\alpha_{vib}^0 = (1/3)\, p_{vib}^2 / |\Delta E|, \qquad (4)$$

termed *vibronic polarizability*, with

$$p_{vib} = p_{12} \sqrt{[1 - (E_{gap}/4E_{JT})^2]} \tag{5}$$

standing for the *vibronic electric dipole*. The vibronic mixing effects give rise to a *vibronic VdW pairing energy* in lieu of equation (2) [4,6]:

$$U_{VdWvib} = \tfrac{1}{2} \Delta E \, (\alpha_{vib}^0 / \kappa R_{ij}^3)^2 = (p_{12}^4/18\kappa^2 \Delta E)(1/R_{ij}^6)[1 - (E_{gap}/4E_{JT})^2]^2 \tag{6}$$

(See References [4,6,7] for more details.) The vibronic polarizability $\alpha_{vib}$ is temperature-dependent ($\alpha_{vib}^0$ is its low-temperature value).. For a molecular system [8]:

$$\alpha_{vib}(T) = \alpha_{vib}^0 \tanh(|\Delta E|/k_B T) \tag{7}$$

From (6) we obtain the attractive part of the electrostatic *VdW binding energy* $V_{vib}$ of a *vibronic two-level system* at 0 K:

$$V_{vib}^0 = \tfrac{1}{2} \Delta E \, (\alpha_{vib}^0/\kappa)^2 \sum_{ij} R_{ij}^{-6} \tag{8}$$

The vibronic energy (8) is to be compared with the attractive part of the *VdW binding energy* $V_{el}$ *of an electronic system in the uncoupled two-level state*:

$$V_{el} = \tfrac{1}{2} E_{gap}(\alpha_{exc}/\kappa)^2 \sum R_{ij}^{-6} \tag{9}$$

Taking the ratio of (8) to (9) we get

$$V_{vib}^0 / V_{el} = [1 - (E_{gap}/4E_{JT})^2]^2 \exp(2E_{JT}/\eta\omega) \tag{10}$$

It is the large exponential term (reciprocal Holstein reduction factor) that makes $V_{vib} / V_{el} \gg 1$. We also note that for small polarons $4E_{JT} \gg E_{gap}$ Equation (10) justifies terming the vibronic VdW interaction energy '*colossal*'.

Equations (9) and (8) for the binding energy of a two-level system may be translated into a binding energy of a small-polaron superlattice on introducing the respective pairing energies (2) and (6), provided each pair is counted once. Assuming additivity of the VdW interaction energy this gives

$$V_{vib}^0 = \tfrac{1}{2}\Delta E \, (\alpha_{vib}^0/\kappa)^2 (\alpha_M/R^6) \tag{11}$$

where $\alpha_M = \sum_{ij} (R/R_{ij})^6$ is the lattice sum constant of the small-polaron superlattice. In as much as the Mn sublattice is simple cubic crystallographically, so is the small-polaron superlattice with a lattice constant R in equation (11). Performing the summation we get $\alpha_M = 6(1 + \tfrac{1}{4} + 1/8 + \ldots)$.

## 2.2. Field coupling to lattice oscillators

We refer the reader to earlier publications for a survey of the coupling of lattice oscillators to an external field [6,7]. The relevant analysis leads to the following field coupling terms: $H'_E = -\mathbf{p}\cdot\mathbf{E}$ and $H'_H = -\boldsymbol{\mu}\cdot\mathbf{H}$ where $\mathbf{p}$ and $\boldsymbol{\mu}$ are the electric and magnetic dipoles, respectively, pertaining to a single oscillator. Earlier analyses considered the environment coupled to a small polaron as a traditional system of linear harmonic oscillators [9] though later improved approaches introduced nonlinear oscillators too [10]. Accordingly $p = \pm p_{gu}\sqrt{[1- (E_{gu}/4E_{JT})^2]}$ which obtains at $p_{gu}\times 2GQ/\sqrt{[(2GQ)^2 + E_{gu}^2]}$ for $Q = \pm Q_{min} = \pm\sqrt{[4G^4 - K^2 E_{gu}^2]}/2GK$ from the linear oscillator model [9], and $\mu = (4\pi\mu_0/c)(\pi\rho^2)\Omega_{rot}$ from the nonlinear model [11]. Here $p_{gu}$ is the electric dipole arising from the breakup of inversion symmetry at the small polaron site, Q is the coupled phonon mode coordinate, K is the stiffness, G is the electron-phonon coupling constant, $\mu_0$ is the magnetic permeability of the medium, $\rho$ and $\Omega_{rot}$ are the small polaron orbital radius and rotational frequency, respectively. The last two quantities are inherent to the nonlinear oscillator representing the small polaron. Under these conditions the electric dipole will be holding good at energies closer to the well bottom, while the magnetic dipole will hold true within most of the energy range.

Introducing the external field into the analysis extended so as to account for the field-coupling terms, we see that the main quantity to be affected by the applied field is the vibronic tunneling splitting $\Delta E$ which turns into

$$\Delta E(F) = \sqrt{[\Delta E(0)^2 + H'^2_F]} \tag{12}$$

where $\Delta E(0)$ is the field-off splitting by equation (3) and $H'_F$ is the field-coupling energy. At low fields $[H'_F \ll \Delta E(0)]$ $\Delta E(F) = \Delta E(0)\{1 + \frac{1}{2} [H'_F/\Delta E(0)]^2\}$, at high fields $[H'_F \gg \Delta E(0)]$ we have $\Delta E(F) = H'_F \{1 + \frac{1}{2} [\Delta E(0)/ H'_F]^2\}$. The field effect on the small-polaron superlattice converts the low-field condition to a high-field one. At any rate, the field-dependent tunneling splitting is superior to the field-off one: Consequently, from equation (6) the field-off pairing energy (6) is superior to the field-on one which implies that the field effect always tends to suppress the conductivity.

## 3. Discussion

After completing most of this manuscript, we downloaded a copy of a very recent experimental report on the electroresistance effect in manganites [12]: Three months ago, a German-American joint group published highly informative results confirming the formation of a small-polaron superlattice above Curie's temperature in $Pr_{0.6}Ca_{0.4}MnO_3$ and interpreted the electroresistance effect as a field-induced order-disorder (solid→liquid) transition. They watched what happens to the sample and its electric resistance on applying a sufficiently high electric field onto it. For that purpose the authors used the tip of a scanning tunneling microscope to apply a field to a small region of the sample and to measure its resistance. The electronic microscope was employed to watching the sample by electron-diffraction which confirmed the solid to liquid conversion when the tip-applied field was sufficiently high.

To get a better idea just how high the field should have been, we make use of eq. (11) and substitute $\Delta E(F)$ for $\Delta E$ to obtain for the lattice energy $V_{vib}^F = (p_{12}^2/\kappa)^2(\alpha_M/R^6)/18\Delta E(F)$. The lattice will melt down when $V_{vib}^F \leq k_BT$ which gives $\Delta E(F) \geq (p_{12}^2/\kappa)^2(\alpha_M/R^6)/18k_BT$. In field terms this leads to $H'_F\{1 + \frac{1}{2}[\Delta E(0)/H'_F]^2\} \geq (p_{12}^2/\kappa)^2(\alpha_M/R^6)/18k_BT$ (high fields) or to $\Delta E(0)\{1 + \frac{1}{2}[H'_F/\Delta E(0)]^2\} \geq (p_{12}^2/\kappa)^2(\alpha_M/R^6)/18k_BT$ (low fields), as set above in Section 2.2. For a crude estimate, we choose the high-field criterion yielding $F \geq (p_{12}^3/\kappa^2)(\alpha_M/R^6)/18k_BT$ (CER) or $H \geq (p_{12}^4/\kappa^2)(\alpha_M/R^6)/18\mu k_BT$ (CMR).

For an order of magnitude estimate, we tentatively set $p_{12} = 1$ eÅ, $\kappa = 5$, $\alpha_M = 1$, $R = 12$ Å, $T = 100$ K. The CER crude criterion yields $F_c = 1.3 \times 10^3$ V/cm $= 1.3 \times 10^5$ V/m (100K) $= 0.4 \times 10^5$ V/m (300K). To compare with, we find an experimental value of $F_c = 0.5 \times 10^5$ V/m from Figure 5 of Ref. [12]. Clearly, our crude criterion may be applicable though by using the complete formulas. For the time being, however, we only aim at drawing attention.

We stress finally that our conjecture for a colossal VdW coupling of a specific form [13] as in Section 2.1 relies essentially on a 2$^{nd}$ order perturbation result as in equation (2) which may not be correct at high fields. For such fields the electrostatic response may rather be determined by the nonlinear polarizability which requires revising the basic equations. This is a challenge for the future though it is not expected to bring in too drastic qualitative changes.

Acknowledgements

I thank Dr. David M. Eagles (London) for his comments on the intermolecular pairing at higher binding energies which may require introducing corrections to the 2$^{nd}$ order perturbation result. I am also grateful to Maria Mladenova for a critical reading.

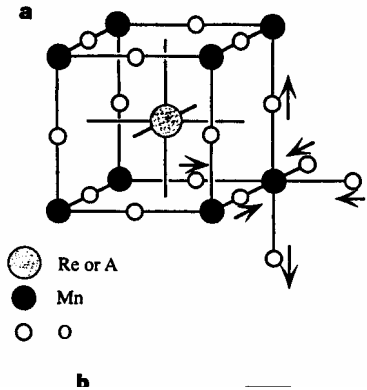

Figure 1(a): The Manganite unit cell. The big ball at the body center is a rare-earth ion, usually $La^{3+}$. The magnetic $Mn^{3+}$ ions are situated at the vertexes of the cube, the compensating $O^{2-}$ ions are edge centered. Each manganese ion is at the center of an $6O^{2-}$ oxygen octahedron whose ions vibrate in an even parity $E_g$ – like mode. The compound formula is $RE_{1-x}A_xMnO_4$ where RE is a rare earth, A is a divalent substitutional cation dopant introduced to implant holes.

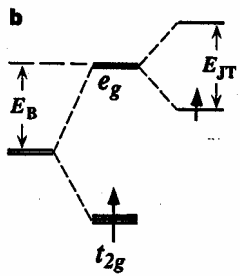

Figure 1(b): The manganese energy levels and their splitting. The original Mn d-electron quintet splits into a $t_{2g}$ triplet and a $e_g$ doublet in the crystalline field. The doublet, if singly occupied, is further split through coupling to the $E_g$ Jahn-Teller mode, the splitting amounts to the Jahn-Teller energy. Both diagrams are taken from Millis' paper in Reference [2].

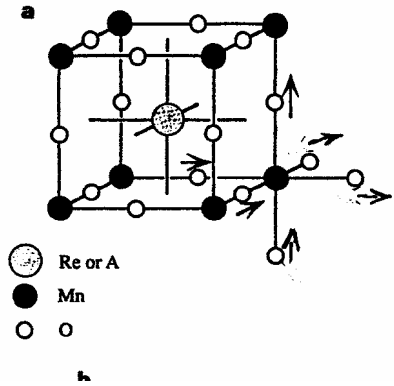

Figure 1(c): Showing the $T_{1u}$- like odd-parity vibration of the oxygen octahedron. Coupling of this mode to a pair of opposite-parity electronic states may result in an off-centering of the oxygen octahedron from the normal Mn site. This will give rise to the breaking down of the inversion site symmetry and the consequent occurrence of an electric dipole. In addition, the octahedron may reorientate through jumping over the off-center sites in an orbital rotation to give rise to orbital magnetic dipoles. The obtained picture apparently contains most of the features needed for the structure to couple to the external fields.

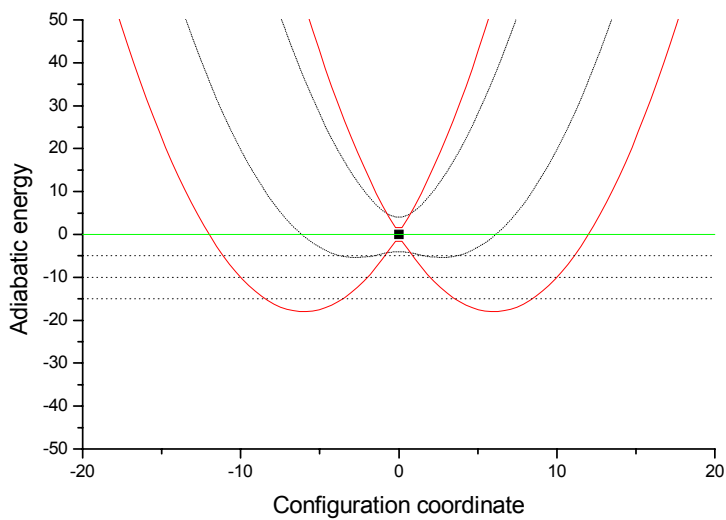

Figure 2: Pairs of adiabatic potentials for small polarons (solid lines) and large polarons (dotted lines). The small polarons are confined locally to very narrow squeezed bands (dotted) while the large polarons are only weakly bound moving in wide bands through the diagram center (solid reference).

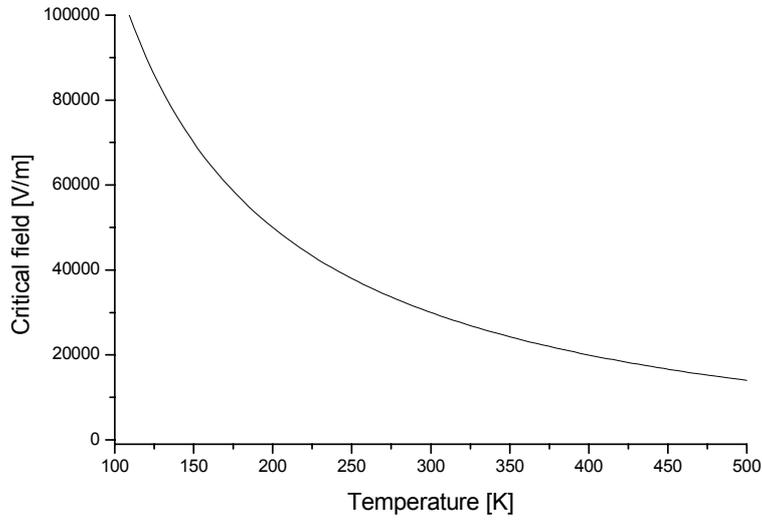

Figure 3: The temperature dependence of the critical electric field from $F_c = (p_{12}^3/\kappa^2) \times (\alpha_M/R^6)/18k_BT - F_\infty$, where the symbols are explained in Sections 2.2 and 3, while the intercept $F_\infty = \Delta E(0)/p_{12}$ is related to the zero-field tunneling splitting. From the available value of $F_\infty \sim 10^4$ V/m at $p_{12} \sim 1$ eÅ, we get $\Delta E(0) \sim 1$ μeV which is not at all an epicyclic estimate. Finally, we use the calculation of Section 3 to get $F_c = 1.20 \times 10^7 / T - 10^4$ [V/m] to calculate a tentative temperature dependence of the critical field.